\documentclass[%
 aip,
 amsmath,amssymb,
 reprint,%
]{revtex4-1}

\usepackage{graphicx}
\usepackage{dcolumn}
\usepackage{bm}

\usepackage[utf8]{inputenc}
\usepackage[T1]{fontenc}
\usepackage{mathptmx}
\usepackage{xcolor}
\usepackage{braket}
\usepackage{array}

\begin{document}

\preprint{AIP/123-QED}

\title[Temporal Quantum Noise Reduction Acquired by an Electron-Multiplying Charge-Coupled-Device Camera]{Temporal Quantum Noise Reduction Acquired by an Electron-Multiplying Charge-Coupled-Device Camera}

\author{Fu Li}%
\affiliation{ 
Department of Physics and Astronomy, Texas A\&M University, College Station, TX 77843
}%
\affiliation{%
Institute for Quantum Science and Engineering,  Texas A\&M University, College Station, TX 77843
}%

\author{Tian Li}
\email{tian.li@tamu.edu}
\homepage{T. Li and F. Li contributed equally to this work.}

\affiliation{%
Institute for Quantum Science and Engineering,  Texas A\&M University, College Station, TX 77843
}%
\affiliation{ 
Department of Biological and Agricultural Engineering, Texas A\&M University, College Station, TX 77843
}%

%

\author{Girish S. Agarwal}
\affiliation{ 
Department of Physics and Astronomy, Texas A\&M University, College Station, TX 77843
}%
\affiliation{%
Institute for Quantum Science and Engineering,  Texas A\&M University, College Station, TX 77843
}%
\affiliation{ 
Department of Biological and Agricultural Engineering, Texas A\&M University, College Station, TX 77843
}%

\date{\today}

\begin{abstract}
Electron-multiplying charge-coupled-device cameras (EMCCDs) have been used to observe quantum noise reductions in beams of light in the transverse spatial degree of freedom. For the quantum noise reduction in the temporal domain, `bucket detectors,' usually composed of photodiodes with operational amplifiers, are used to register the intensity fluctuations in beams of light within the detectors' bandwidth. Spatial information, however, is inevitably washed off by the detector.  In this paper, we report on measurements of the temporal quantum noise reduction in bright twin beams using an EMCCD camera. The four-wave mixing process in an atomic  rubidium vapor cell is used to generate the bright twin beams of light.  We observe $\sim$~25\% of temporal quantum noise reduction with respect to the shot-noise limit in images captured by the EMCCD camera. Compared with bucket detectors, EMCCD makes it possible to take advantage of the spatial and
temporal quantum properties of light simultaneously, which would greatly benefit many applications using quantum protocols.
\end{abstract}

\maketitle

\section{Introduction}

Quantum noise fluctuations in a beam of light below the shot-noise limit (SNL), i.e., squeezed light, was first observed in a groundbreaking experiment by Slusher \textit{et al.}~\cite{PhysRevLett.55.2409}
Since then squeezed light was implemented to enable enhanced communication rates~\cite{PhysRevLett.90.167903,1055958,1056033,PhysRevA.61.042302} and improved detection of weak forces such as gravitational waves~\cite{PhysRevD.23.1693,Chua_2014}. The latter was demonstrated first at the GEO600 gravitational wave detector~\cite{Abadie:2011bs} and later at the LIGO detector~\cite{Aasi:2013fv}. These applications, although proposed more than three decades ago, are still some of the most prominent applications of squeezed light. In addition to these applications, squeezed states have also been shown to be the resource of quantum teleportation~\cite{PhysRevLett.80.869,Furusawa706}, continuous-variable quantum computing~\cite{PhysRevLett.97.110501}, quantum error correction coding~\cite{Aoki:2009dz,Lassen:2010fu}, phase estimation~\cite{Berni:2015kl} and tracking~\cite{Yonezawa1514}, fundamental tests of quantum mechanics (such as the Einstein-Podolsky-Rosen gedanken experiment)~\cite{PhysRevLett.60.2731,RevModPhys.81.1727,PhysRevLett.68.3663}, quantum imaging~\cite{PhysRevLett.85.3789,PhysRevLett.88.203601} of e.g., biological samples~\cite{Taylor:2013qa}, clock synchronization~\cite{Giovannetti:2001mi} and magnetometry~\cite{PhysRevLett.105.053601,Otterstrom:14}. Moreover in recent years, a squeezed light source has been the working horse for quantum state engineering, in particular non-Gaussian state generation using the method of photon subtraction~\cite{Ourjoumtsev83,PhysRevLett.97.083604,DELLANNO200653,Andersen:2015pi} as required for various quantum processing protocols~\cite{PhysRevLett.89.137903,PhysRevLett.89.137904,PhysRevA.66.032316,PhysRevLett.102.120501,PhysRevLett.88.097904}. 

It is interesting to note that the experimental platforms for generating squeezed light, such as nonlinear crystals, fibers and atomic ensembles used in the 80's are still the same as those used today for generating much more efficient squeezing. Although significant advancements have been made from the initial 0.3~dB squeezing~\cite{PhysRevLett.55.2409} till today's near 15~dB squeezing~\cite{PhysRevLett.117.110801}, those advancements have mainly been of technical nature, i.e., successful development of low-noise electronics for phase locking, low loss optical components and high quantum efficiency photodiodes have led to largely improved systems.  

Most of the aforementioned studies pertaining to squeezed light are in the temporal domain acquired by `bucket detectors', i.e., photodiodes with operational amplifiers having sufficient bandwidth. Nevertheless, squeezed light can also be achieved in the transverse spatial degree of freedom using electron-multiplying charge-coupled-device cameras (EMCCDs)~\cite{Samantaray:2017ff,Brida:2010lh,PhysRevA.95.053849,PhysRevA.100.063828}. In this paper, we report yet another important technical advance in the squeezed light arena, which allows one to use an EMCCD camera to achieve squeezing measurements in the \textit{temporal} domain as well. This makes it possible to take advantage of the temporal and spatial quantum properties of light simultaneously. Many aspects in quantum optics, especially quantum metrology and quantum imaging, could greatly benefit from the concurrent measurement of the quantum correlations in both the temporal and the spatial domains~\cite{Treps940,Genovese_2016,QuantumImaging}.

\section{Experimental setup} 

\begin{figure}
    \begin{center}
    \includegraphics[width=1.02\columnwidth]{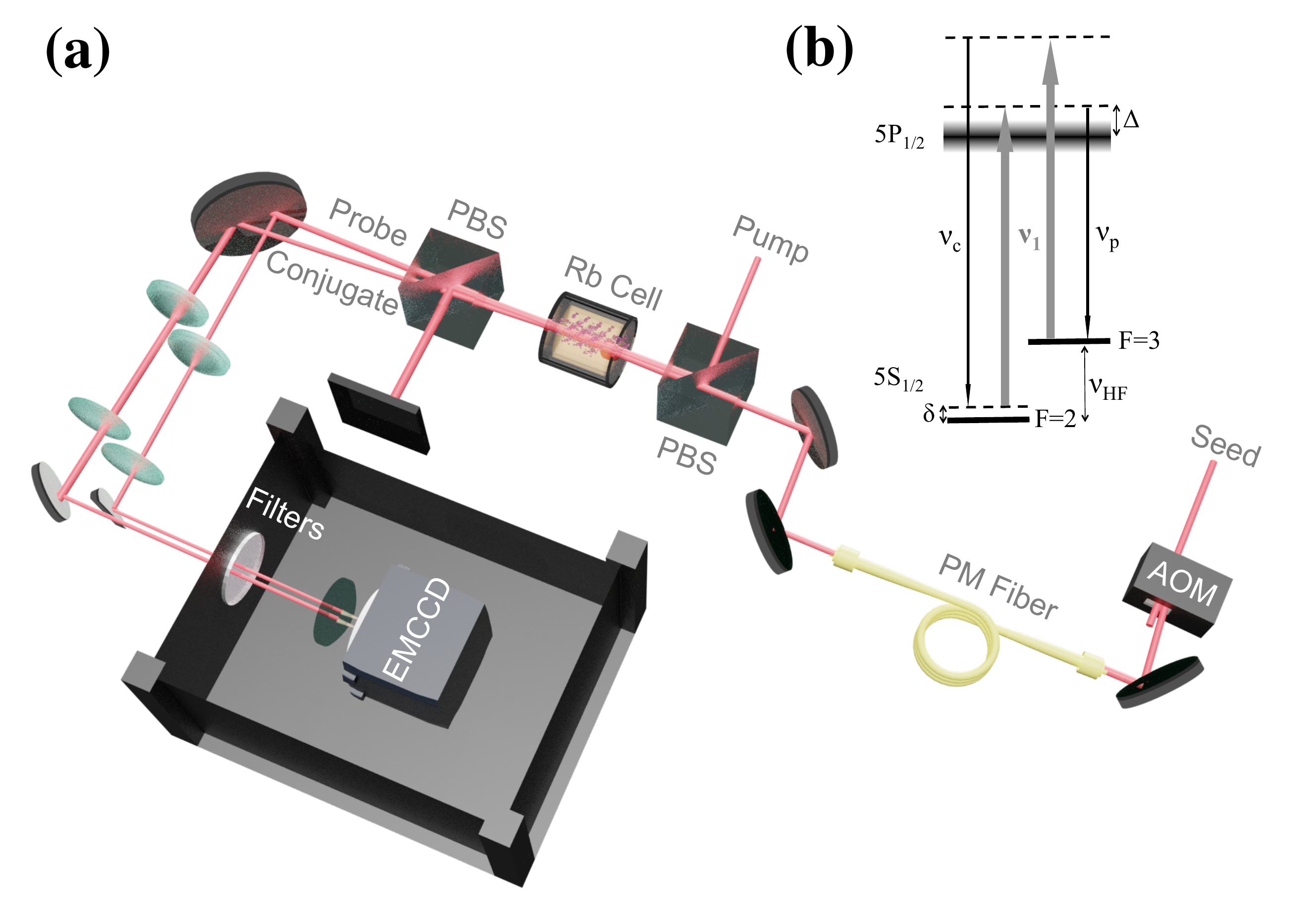}
    \caption{(a) Experimental setup in which a seeded $^{85}$Rb vapor cell produces strong quantum-correlated twin beams via FWM. The twin beams are separated from the pump by a $2\times10^5:1$ polarizer and then focused onto the EMCCD camera. The camera is enclosed in a light-proof box with filters mounted to block ambient light. The AOM on the probe beam path is used to pulse the twin beams with 1~$\mu$s FWHM and duty cycle of $1/12$. PBS: polarizing beam splitter, PM filer: polarization-maintaining fiber. (b) Level structure of the D1 transition of $^{85}$Rb atom. The optical transitions are arranged in a double-$\Lambda$ configuration, where $\nu_p$, $\nu_c$ and $\nu_1$ stand for probe, conjugate and pump frequencies, respectively, fulfilling $\nu_p$ +  $\nu_c$ =  $2\nu_1$. The width of the excited state in the level diagram represents the Doppler broadened line. $\Delta$ is the one-photon detuning, $\delta$ is the two-photon detuning, and $\nu_{\text{HF}}$ is the hyperfine splitting in the electronic ground state of $^{85}$Rb.} 
		\label{Setup}
    \end{center}
\end{figure}

The squeezed light generated in this work is based on the four-wave mixing (FWM) process in an atomic $^{85}$Rb vapor cell~\cite{Dowran:18,PhysRevA.95.023803,Hudelist:2014db,Anderson:17,Li:17,Pooser:15,Clark:2014vf}. The experimental setup and the respective atomic level structure are shown in Fig.~\ref{Setup}(a) and (b). The medium possesses a large third-order electric susceptibility $\chi^{(3)}$ and is pumped by a strong ($\sim 500$~mW) narrow-band continuous-wave (CW) laser at frequency $\nu_1$ ($\lambda = 795$~nm) with a typical linewidth $\Delta \nu_1 \sim 100$~kHz. Applying an additional weak ($\sim$~\textcolor{black}{10}~nW) coherent seed beam 
at frequency $\nu_p = \nu_1 - (\nu_{HF}+\delta)$, where $\nu_{HF}$ and $\delta$ are the hyperfine splitting in the electronic ground state of $^{85}$Rb and the two-photon detuning respectively in Fig.~\ref{Setup}(b) (further experimental details can be found in Ref.~\onlinecite{2phLi}), two pump photons are converted into a pair of twin photons, namely `probe $\nu_p$' and `conjugate $\nu_c$' photons, adhering to the energy conservation $2 \nu_1 = \nu_p + \nu_c$ (see the level structure in Fig.~\ref{Setup}(b)). The resulting `bright twin beams' are strongly quantum-correlated and are also referred to as (seeded) two-mode squeezed light~\cite{PhysRevA.78.043816}.

\begin{figure}
    \begin{center}
    \includegraphics[width=0.9\columnwidth]{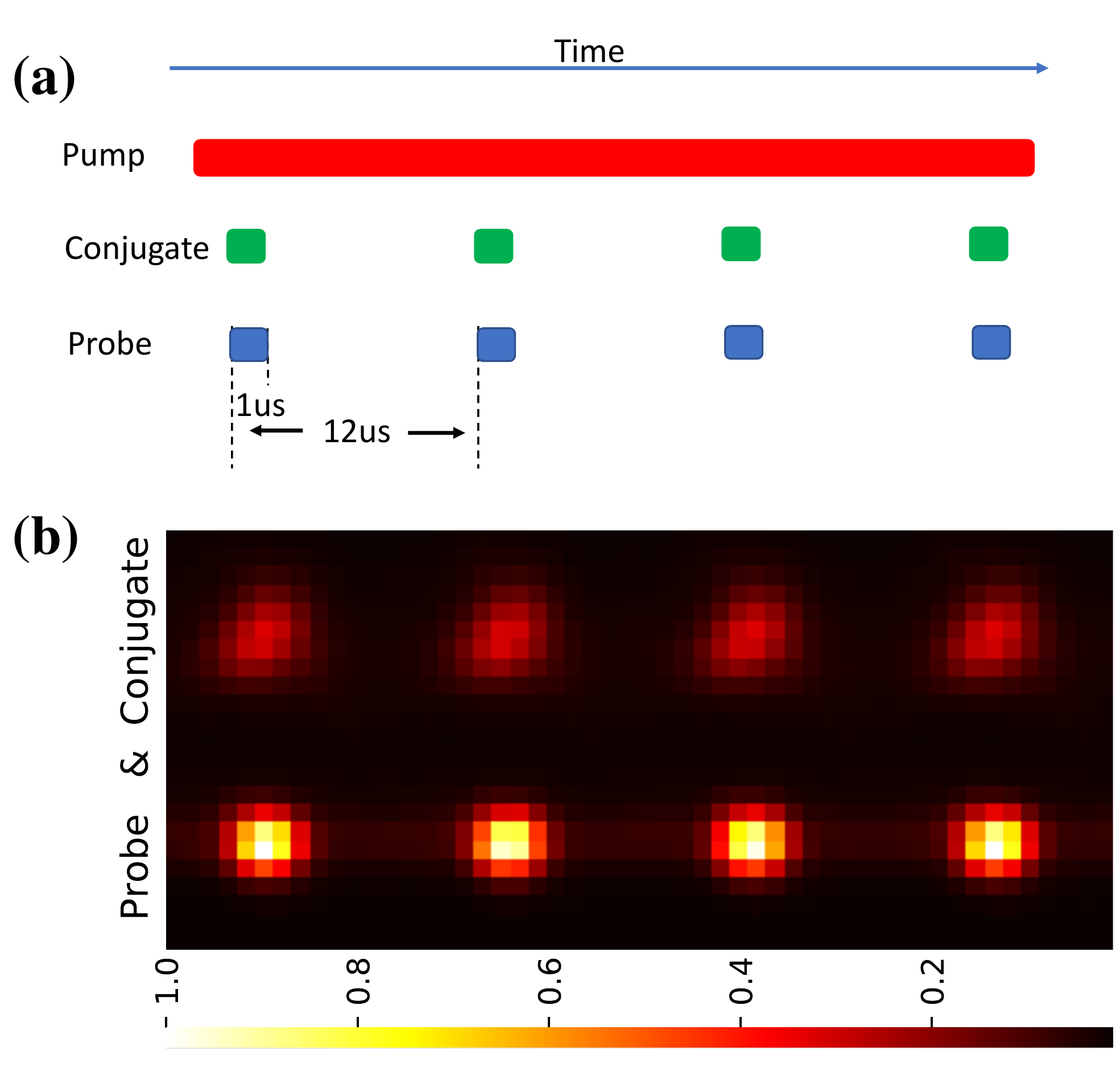}
    \caption{
        \label{Method}
    %
    %
   (a) Time sequencing of the pump and twin beams. The pulse duration of 1~$\mu$s and the duty cycle of 1/12 is realized by pulsing the probe beam with an AOM. The CW pump beam is present all the time. (b) Typical images of the twin beams captured by the EMCCD camera with four consecutive pulses. 
}
    \end{center}
\end{figure}

After the $^{85}$Rb vapor cell,  the pump and the bright twin beams are separated by a second polarizer, with $\sim$~$2\times10^5:1$ extinction ratio for the pump. The twin beams are then focused onto an EMCCD camera (Andor iXon Ultra 897). The EMCCD camera is enclosed in a light-proof box with filters installed at the entrance to block ambient light photons from entering the camera. The acousto-optic modulator (AOM) on the probe beam path is used to pulse the beam with 1~$\mu$s duration (FWHM) and duty cycle of $1/12$. Since the CW pump beam is present all the time, the conjugate beam is therefore also pulsed as a result of the FWM process.  The time sequencing of the pump and the twin beams are shown in Fig.~\ref{Method}(a) as the red strap, and the blue and green squares respectively.  


\section{Results \& Analysis} 

\subsection{Temporal two-mode squeezing measured by photodiodes}

\begin{figure}
    \begin{center}
    \includegraphics[width=1\columnwidth]{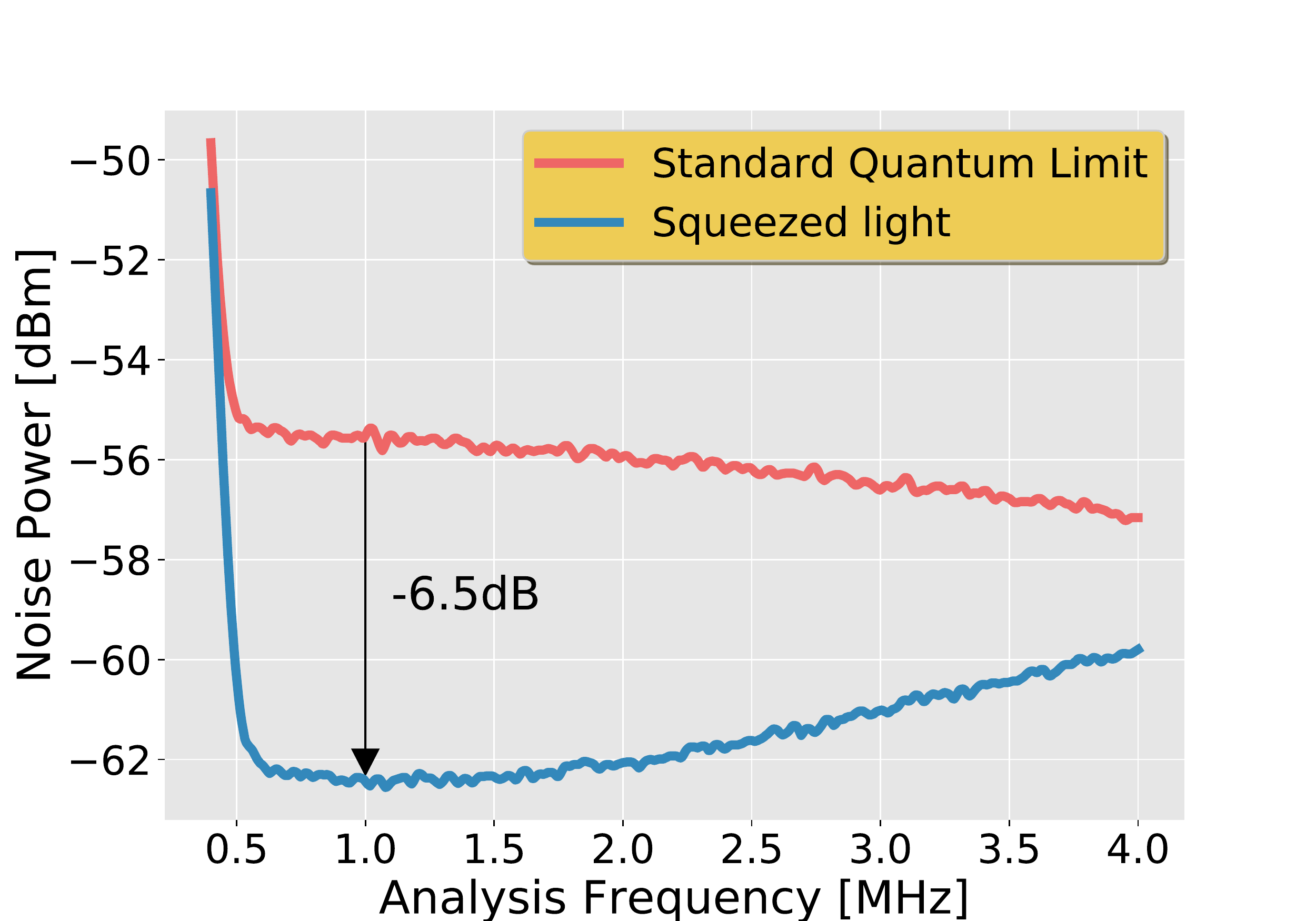}
    \caption{
        \label{SQZ}
    %
    %
  Intensity-difference noise power spectrum for the squeezed twin beams (blue line) and for the standard quantum limit (red line), obtained by a balanced photodetector in conjunction with a radio frequency spectrum analyzer (with a resolution and video frequency bandwidth of 300~kHz and 100~Hz, respectively). A two-mode squeezing of 6.5~dB is achieved around the analysis frequency of 1~MHz.
}
    \end{center}
\end{figure}

We first measure the two-mode squeezing in a conventional way, i.e., using photodiodes to register intensity fluctuations in the beams of light in the temporal domain. After the second polarizer, we direct the probe and conjugate beams into the two ports of a balanced, amplified photodetector with a transimpedance gain of $10^5$ V/A and 94\% quantum efficiency at $\lambda = 795$~nm (not shown in Fig.~\ref{Setup}(a)). The photodetector signals are sent to a radio frequency spectrum analyzer with a resolution bandwidth RBW of 300~kHz and a video frequency bandwidth VBW of 100~Hz. A typical squeezing spectrum is shown in Fig.~\ref{SQZ} as the blue curve. The standard quantum limit (red curve) of this system is measured by picking off the probe before the cell, splitting it with a 50/50 non-polarizing beam splitter, and directing the resulting beams into the balanced, amplified photodetector. The balanced detection technique subtracts away common-mode noise to better than 25~dB. The balanced photodetector noise level is a measure of the standard quantum limit for the total amount of optical power arriving at the photodetector. The standard quantum limit should be independent of frequency, which is indeed the case within the bandwidth of the detection electronics, which begins to drop down above~3 MHz. We measure more than 6~dB of the two-mode squeezing around the analysis frequency of 1~MHz.

\subsection{Temporal quantum noise reduction acquired by an EMCCD camera}

We acquire the temporal quantum noise reduction of the twin beams through the use of the \textit{kinetic} mode of the EMCCD camera. The EMCCD has $512\times512$ pixels with each pixel size of 16~$\mu$m$\times$16~$\mu$m. We focus the twin beams on the camera with an $1/e^2$ beam diameter of $\sim 50~\mu$m, occupying roughly 3 pixels as shown in Fig.~\ref{Method}(b). The temperature of the EMCCD is  kept at $-75^\circ$C to curb the thermal noise contributions. 

Since the pulse duration is 1~$\mu$s and the time interval between two consecutive pulses is 12~$\mu$s, thus in order to completely transfer all charges from the camera's image area to the storage area within one pulse cycle, we can in principle choose to set the speed of vertical pixel shift (i.e., the time taken to vertically shift all pixels one row down) to any value as long as it is faster than 4~$\mu$s, given our beam size is merely 3 pixels across. However, the drawback with a fast vertical pixel shift speed is the reduction of charge transfer efficiency, which in turn causes `vertical smearing' (i.e., light is still falling on the image area during the short time taken to transfer the charge from the image area to the storage area). In our case, we found a 0.9~$\mu$s vertical pixel shift speed in conjunction with a vertical clock voltage amplitude of 4 (to ensure that extremely high signals can be fully removed during the EMCCD clean cycle) worked best for us. 

Another important setting of the EMCCD is the readout rate. It also ought to be fast enough to be within one pulse cycle. However, a faster readout rate always results in a higher readout noise. In our case, we adopt 3~MHz as our readout rate although technically it can be as fast as 17~MHz, but the price one has to pay is 8 fold more readout noise. 

For each measurement, we capture 200 kinetic series (i.e., 200 frame sequences), with each frame containing 35 pairs of probe and conjugate images. For the measurement of the quantum noise reduction, we adopt a similar algorithm developed in Refs.~\onlinecite{PhysRevA.95.053849,PhysRevA.100.063828} but implement it in the temporal domain. In brief we crop a \textcolor{black}{$10\times$10} pixel region around the maximum-intensity region in each probe and conjugate images, large enough to enclose their respective full beam profiles (see Fig.~\ref{Method}(b)), we then are able to obtain the temporal photon counts fluctuations of the probe $N_p(t)$ and conjugate $N_c(t)$ by integrating the photon counts in the copped regions.  The redefined quantum noise reduction characterization~\cite{PhysRevA.95.053849,PhysRevA.100.063828}, $\sigma$, in the temporal domain reads

\begin{figure}[]
    \begin{center}
    \includegraphics[width =1\columnwidth]{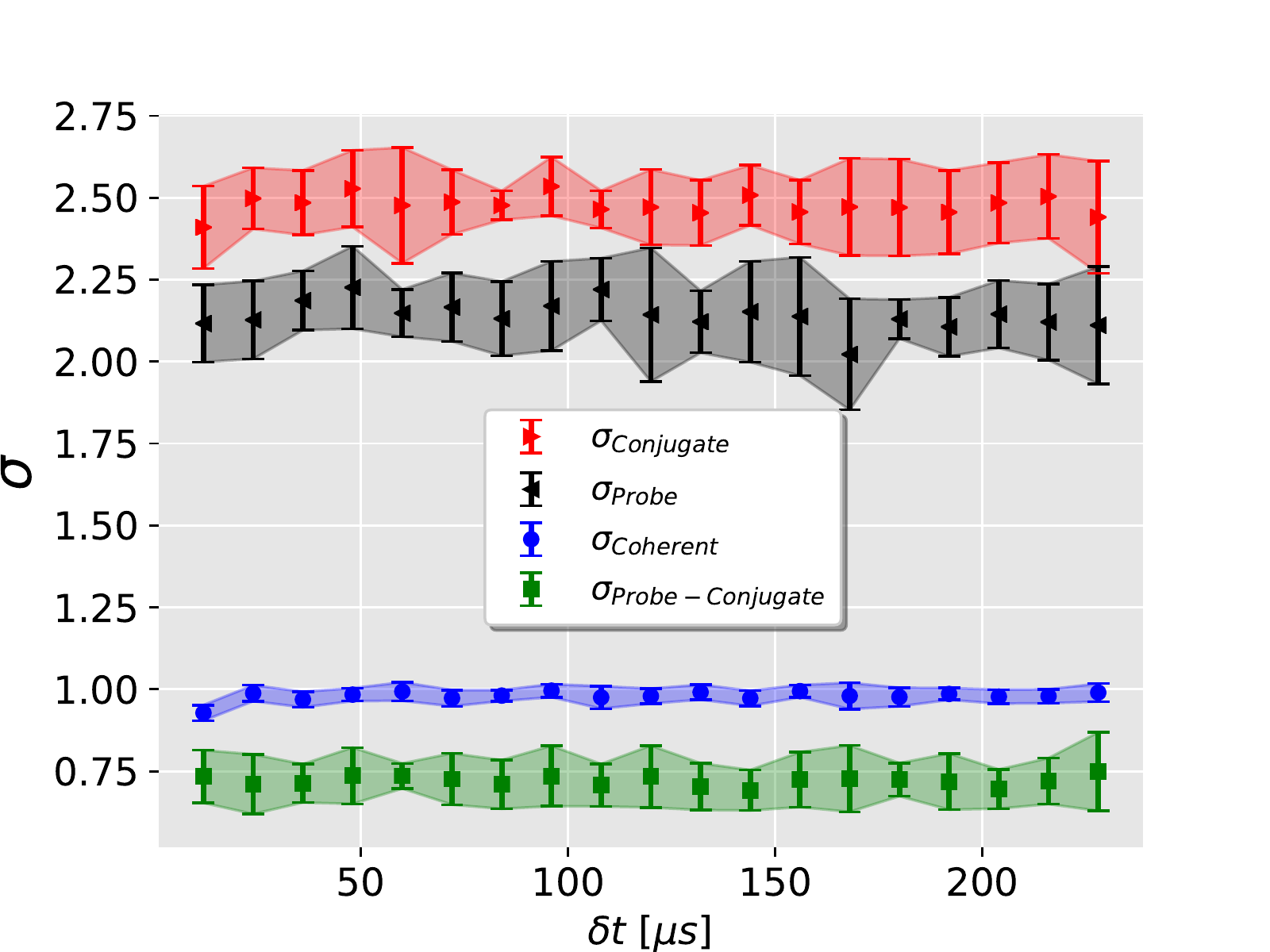}
    \caption{
        \label{Result}
    %
    %
 The temporal quantum noise reduction characterization $\sigma$ as a function of the time interval $\delta t$ between successive images for different beams of light. Green squares: twin beams; blue dots: coherent beams; black triangles: probe beam only; red triangles: conjugate beam only.
        }
    \end{center}
\end{figure}

\begin{equation}
\begin{aligned}
\sigma\equiv\frac{\langle\Delta^2[(N_p(t+\delta t)-N_p(t))- (N_c(t+\delta t)-N_c(t))]\rangle_t}{\langle N_p(t+\delta t)+N_p(t) + N_c(t+\delta t) + N_c(t)\rangle_t}, 
\end{aligned}
\label{sigma}
\end{equation}
where $N_p(t+\delta t)-N_p(t)$ and $N_c(t+\delta t)-N_c(t)$ are the subtractions of photon counts in the cropped regions in two successive probe and conjugate images with time interval of $\delta t$. The subtraction of the two successive images leads to the cancellation of the low-frequency portion of the classical noise as well as the Gaussian profiles of the probe and conjugate images~\cite{PhysRevA.95.053849,PhysRevA.100.063828}. The numerator of Eq.~(\ref{sigma}) represents the temporal variance of the intensity-difference noise between the probe and conjugate pulses. The denominator gives the mean photon counts for the probe and conjugate pulses used for the analysis and represents the shot noise. For coherent state pulses $\sigma  = 1$, which corresponds to the shot noise limit, while for thermal light or other classical states $\sigma >1$.  Temporally quantum correlated beams, like the twin beams generated in our experiment, will result in $\sigma <1$, with a smaller $\sigma$ corresponding to a larger degree of two-mode squeezing.

In Fig.~\ref{Result}, we plot $\sigma$ as a function of $\delta t$ for different beams of light. For each $\delta t$, we average 5 sets of 200 kinetic series and designate the error bar with one standard deviation. As expected, $\sigma \doteq 0.75 < 1$ for the twin beams (green squares), while $\sigma =1$ when the twin beams are replaced with two coherent beams (blue dots), and $\sigma >1$ for the probe beam (black triangles) and conjugate beam (red triangles) individually~\cite{McCormick:07} (calculated as $\sigma_{p,c}=\langle\Delta^2[N_{p,c}(t+\delta t)-N_{p,c}(t)]\rangle_t /\langle N_{p,c}(t+\delta t)+N_{p,c}(t)\rangle_t$). The notable degradation of the temporal quantum noise reduction measured by the EMCCD camera with respect to the one measured by the balanced photodiodes in Fig.~\ref{SQZ} can be mainly attributed to a much worse common noise rejection (caused by the significant mismatch between the spatial modes of the twin beams as shown in Fig.~\ref{Method}(b)) and a much worse quantum efficiency of the EMCCD at 795~nm (less than 80\%  as compared to 94\% for the photodiodes).
 We also repeated the experiment with different pulse duty cycles, but they seemed to play an nonessential role on the quantum noise reduction as long as we were in the shot-noise-limited regime, i.e., the $\sigma$ is still close to unity for coherent beams.

It is worthy to mention that our quantum noise reduction  $\sigma \doteq 0.75$ for the twin beams in the temporal domain is similar to the one reported in Refs.~\onlinecite{PhysRevA.95.053849,PhysRevA.100.063828} in the spatial domain. This is also as expected since when $\delta t = 12$~$\mu$s, we recover the `full spatial mode' case in Refs.~\onlinecite{PhysRevA.95.053849,PhysRevA.100.063828}.

\section{Conclusions}

In conclusion, we report a measurement scheme that is capable of acquiring the quantum noise reduction in the temporal domain using an EMCCD camera. We observe $\sim$~25\% of temporal quantum noise reduction with respect to the shot-noise limit in images captured by the camera. To the best of our knowledge, this is the first experimental showcase that an EMCCD camera can be used to acquire quantum properties of light not only in the spatial domain~\cite{Samantaray:2017ff,Brida:2010lh,PhysRevA.95.053849,PhysRevA.100.063828}, but also in the temporal domain as well. 

We use FWM in an atomic $^{85}$Rb vapor cell to generate the quantum-correlated twin beams of light. Major advantages of this quantum light generation scheme are narrow-band probe and conjugate beams ($\sim 20$~MHz)~\cite{Clark:2014vf,Glasser2012a} along with an ultra-high photon-pair flux up to~$10^{16}$ photons/s~\cite{Dowran:18,PhysRevA.95.023803,Hudelist:2014db,Anderson:17,Li:17,Pooser:15,Clark:2014vf}, which is a few orders of magnitude higher than the fluxes produced by spontaneous parametric down-converters (SPDCs).  Therefore the bright twin beams can be readily applied in some atom-light interaction based quantum protocols~\cite{Duan:2001bd}. Moreover, the FWM process offers sufficient gains in a single-pass configuration producing bright quantum-correlated beams of light without a cavity~\cite{PhysRevLett.100.033602}. This makes it possible to preserve the multi-spatial-mode nature of the bright twin beams~\cite{PhysRevLett.109.043602,Corzo:11} and to observe spatial quantum correlations in the macroscopic regime~\cite{PhysRevA.95.053849,PhysRevA.100.063828}. Our quantum light generation together with the measurement scheme reported here can therefore pave the way for many applications in quantum metrology and quantum imaging, which would greatly benefit from the concurrent measurements of the quantum correlations in both the temporal and the spatial domains~\cite{Treps940,Genovese_2016,QuantumImaging}.

\begin{acknowledgments}
This work is supported by the Air Force Office of Scientific Research 
(Award No. FA-9550-18-1-0141) and the Robert A. Welch Foundation (Award No. A-1943-20180324). F. L. acknowledges support from the Herman F. Heep and Minnie Belle Heep Texas A\&M University Endowed Fund held/administered by the Texas A\&M Foundation.
\end{acknowledgments}

\section*{Data Availability Statement}

The data that support the findings of this study are available from the corresponding author upon reasonable request.

\section*{Reference}
\bibliography{MyLibrary}

\end{document}